\documentclass[twocolumn,showpacs,preprintnumbers,amsmath,amssymb,showkeys,pra,10pt]{revtex4-1}
\usepackage{hyperref}
\usepackage{graphicx}% Include figure files
\usepackage{dcolumn}% Align table columns on decimal point
\usepackage{bm}% bold math
\usepackage[T1]{fontenc}
\newcommand{\szerokosc}{0.495}
\newcommand{\szerokoscdwa}{0.495}
\newcommand{\linkDOI}[1]{\href{http://dx.doi.org/#1}{#1}}
\begin{document}

\preprint{Submitted to: ACTA PHYSICA POLONICA A}

\title{Monte Carlo study of phase separation in magnetic insulators}

\author{Szymon Murawski}%
\author{Konrad Jerzy Kapcia}%
    \email[corresponding author; e-mail:]{konrad.kapcia@amu.edu.pl}
\author{Grzegorz Paw\l{}owski}%
\author{Stanis\l{}aw Robaszkiewicz}%
\affiliation{Electron States of Solids Division, Faculty of Physics, Adam Mickiewicz University in Pozna\'n, Umultowska 85, 61-614 Pozna\'n, Poland
}

\date{April 2, 2015}

\begin{abstract}
In this work we focus on the study of phase separation in the zero-bandwidth extended Hubbard with nearest-neighbors intersite Ising-like magnetic interactions $J$ and on-site Coulomb interactions $U$.
The system has been analyzed by means of Monte Carlo simulations (in the grand canonical ensemble) on two dimensional square lattice (with \mbox{$N=L\times L =400$} sites) and the results for \mbox{$U/(4J)=2$} as a function of chemical potential and electron concentration have been obtained.
Depending on the values of interaction parameters the system exhibits homogeneous (anti-)ferromagnetic (AF) or non-ordered (NO) phase as well as phase separation PS:AF/NO state. Transitions between homogeneous phases (i.e. \mbox{AF--NO} transitions) can be of first or second order and the tricritical point is also present on the phase diagrams.
The electron compressibility $K$ is an indicator of the phase separation and that quantity is of particular interest of this paper.
\end{abstract}

%\pacs{71.10.Fd, 75.10.-b, 75.30.Fv, 64.75.Gh, 71.10.Hf}

%Use showkeys class option if keyword
                              %display desired

\maketitle

\section{Introduction}

Magnetic insulators are a class of materials realized in many various compounds.
An example of them are materials known as transition metal cluster compounds with general formula $AM_4X_8$, where $A$---trivalence metal, $M$---transition metal, $X$---chalcogenide \cite{B1973,RBCT1983,LRK1997}.
Phase separations can occur in magnetic insulators in various circumstances and their theoretical understanding is very current topic. Moreover, instabilities such as stripe formation (as well as charge order) can occur in high-$T_c$ superconductors (e.g. in cuprates)~\cite{BFKT2009}.
A simplified model to describe behavior of such materials is the extended Hubbard model with intersite magnetic interactions \cite{MRR1990,JM2000,DJSZ2004,CzR2006,DJSTZ2006,CzR2006a}.
In this work we study the zero-bandwidth limit of the extended Hubbard model. The Hamiltonian of this model has the form:
\begin{equation}\label{row:ham}
\hat{H} = U\sum_i{\hat{n}_{i\uparrow}\hat{n}_{i\downarrow}} + 2J\sum_{\langle i,j\rangle}{\hat{s}^z_{i}\hat{s}^z_{j}} - \mu\sum_{i}{\hat{n}_{i}}
\end{equation}
where $U$~is the on-site density-density interaction,
$J$~is \mbox{$z$-component} of the intersite magnetic exchange interaction, and
$\mu$ is chemical potential.
The interactions are effective model parameters and are assumed to include
all the possible contributions and renormalizations.
$\sum_{\left\langle i,j\right\rangle}$ restricts the summation to nearest neighbors (independently).
\mbox{$\hat{n}_i=\hat{n}_{i\uparrow}+\hat{n}_{i\downarrow}$} is total electron number on site $i$ and
\mbox{$\hat{s}^z_i=(1/2)(\hat{n}_{i\uparrow}-\hat{n}_{i\downarrow})$} is $z$-component of total spin at $i$ site.
\mbox{$\hat{n}_{i\sigma}=\hat{c}^{+}_{i\sigma}\hat{c}_{i\sigma}$} is electron number with spin $\sigma$ on site $i$,
where $\hat{c}^{+}_{i\sigma}$ and $\hat{c}_{i\sigma}$ denote the creation and annihilation operators, respectively, of an electron with spin $\sigma$ (\mbox{$\sigma=\uparrow,\downarrow$}) at the site $i$.
The electron concentration $n$ is defined as \mbox{$n=(1/N)\sum_i \langle \hat{n}_i \rangle$}, where $N$ is the total number of sites.

The  model studied exhibits two symmetries: (i)~the symmetry between \mbox{$J>0$} (antiferromagnetic) and \mbox{$J<0$} (ferromagnetic) cases and (ii) the electron-hole symmetry.  Because of these symmetries only analyses for \mbox{$0\leq n\leq 1$} and \mbox{$J>0$} have been performed.

\begin{figure*}[t]
\centering
\includegraphics[width=\szerokosc\textwidth]{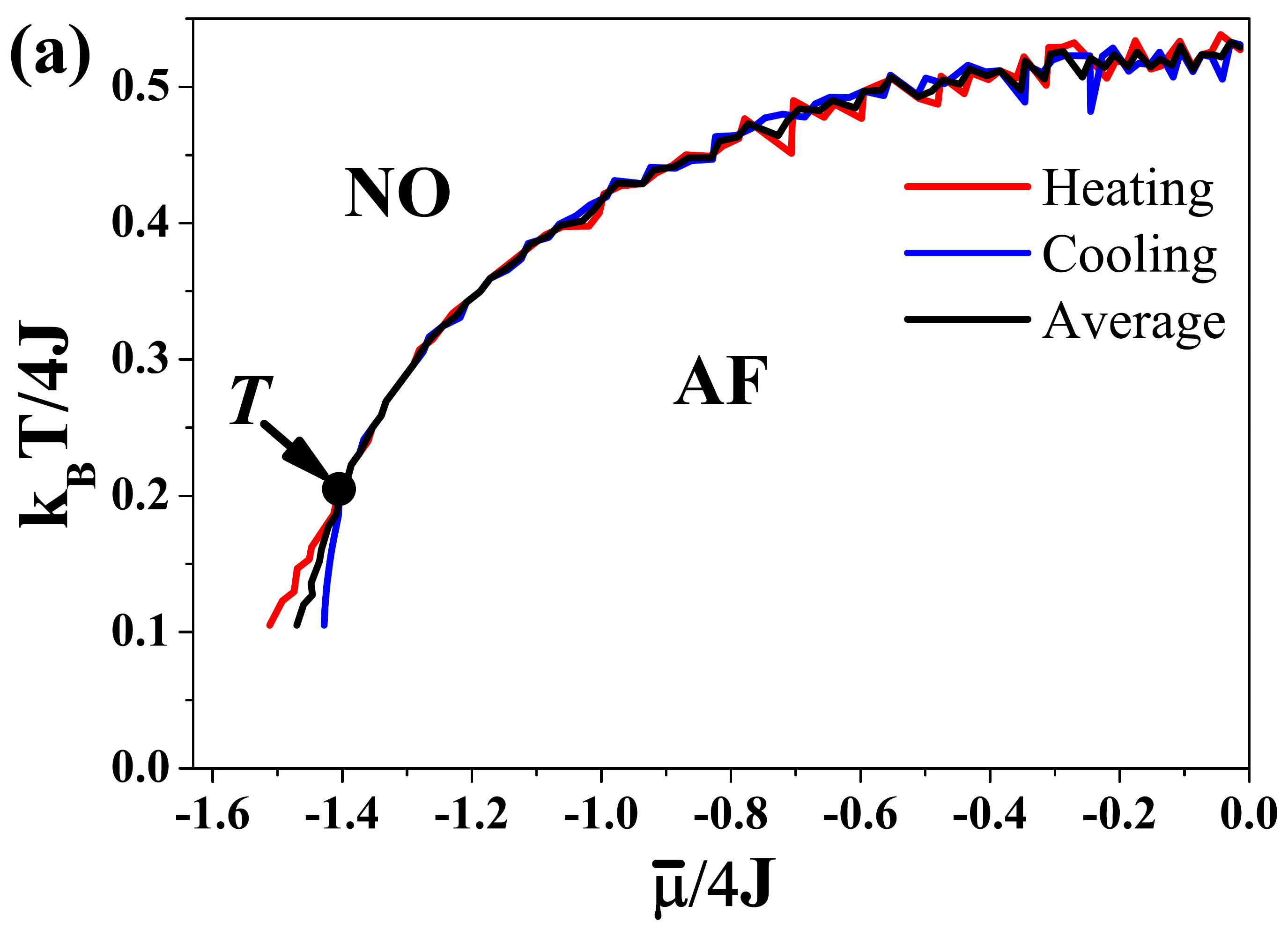}
\includegraphics[width=\szerokosc\textwidth]{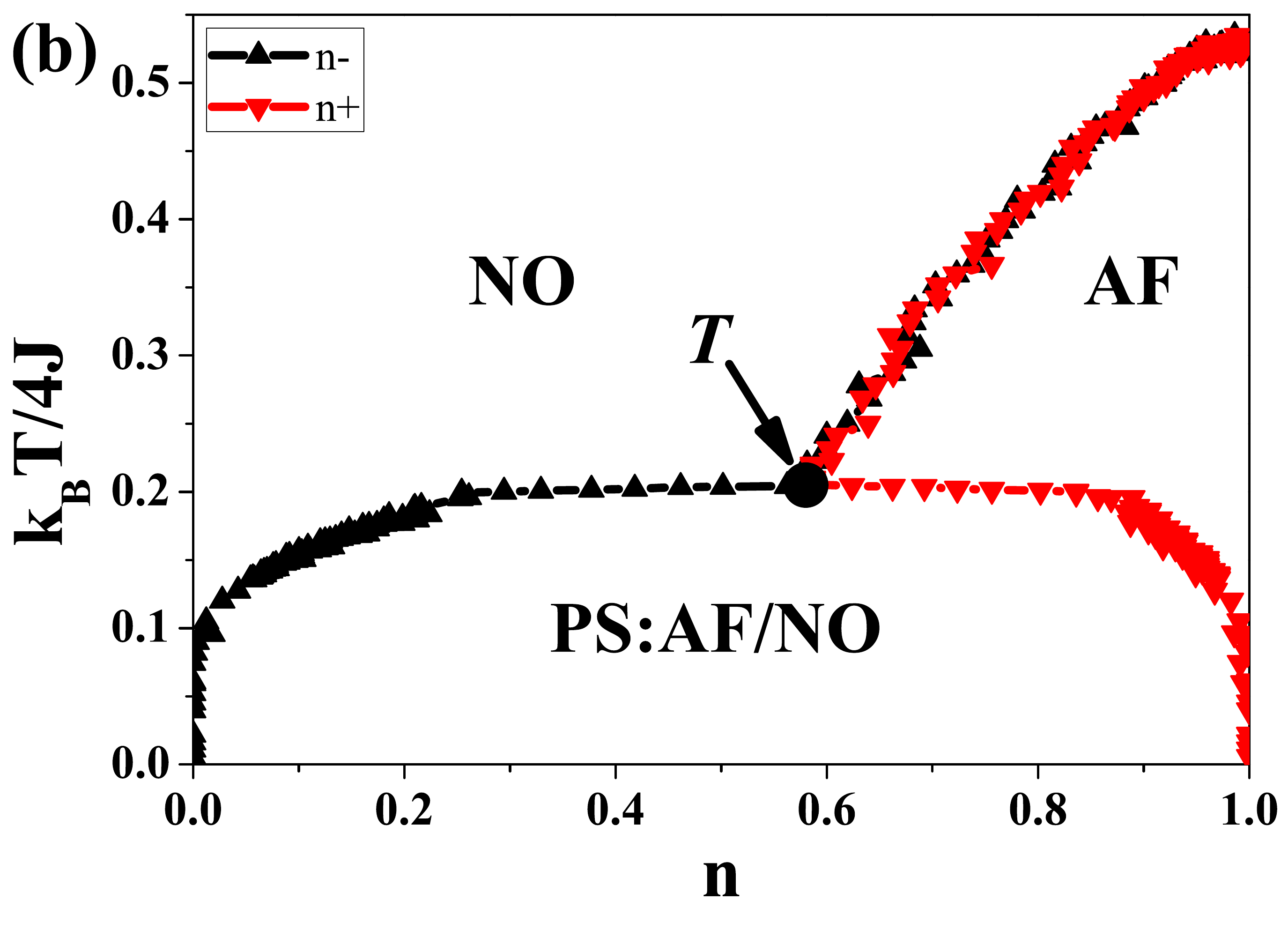}
\caption{The phase diagrams of the model for \mbox{$U/4J=2$}: (a) $k_BT/4J$~vs.~$\bar{\mu}/4J$ and (b) $k_BT/4J$~vs.~$n$ (\mbox{$L=20$}).
$\mathbf{T}$ denotes a tricritical point. On panel (a) ``heating'' and ``cooling'' label the boundaries obtained by simulation performed with increasing and decreasing temperature, whereas ``average'' is the average of these two results. They differ if the \mbox{AF--NO} transition is first-order.}
\label{fig:PhaseDiagrams}
\end{figure*}

We have used the Monte Carlo (MC) simulations to analyze the system. Simulations have been done using Hamiltonian described above on two dimensional square lattice with \mbox{$N=L\times L$} sites in the grand canonical ensemble, which allows us to obtain e.g. chemical potential dependence of electron concentration curves --- $n(\mu)$.
The Monte Carlo algorithm used in this analysis consists of three steps: (i) creation, (ii) destruction, and (iii) moving of particle, all of them with appropriate probability \mbox{$P\sim\exp{\left(\Delta E/(k_BT)\right)}$} \cite{P2006,PK2008,MKPR2012,MKPR2014}.
It is worth noting, that for constant values of concentration a~simpler algorithm with only step (iii) --- ,,move'' would be sufficient.
However, addition of the grand canonical parts (creation and destruction) allows one for more detailed analysis in full range of chemical potential and concentration.
Unfortunately, addition of chemical potential term in the Hamiltonian prevents us from implementing cluster updates algorithm, so only local updates \cite{H1990} are used here.
The details of the algorithm used can be found in \cite{P2006,PK2008,MKPR2012,MKPR2014}.

The exact ground state (\mbox{$T=0$}) results for this model have been found in the case of a~\mbox{$d=1$} chain \cite{MPS2012,MPS2013} using the Green function formalism as well as  for \mbox{$2\leq d< +\infty$} case \cite{BS1986,J1994}.
The rigorous results for finite temperatures \mbox{$T>0$} have been also obtained \cite{MPS2013,N1} for \mbox{$d=1$} chain (an absence of long-range order at \mbox{$T>0$}).
Within the variational approach (with on-site $U$ term treated exactly and mean field decoupling of intersite term $J$)  the model has been analyzed for half-filing (\mbox{$n=1$}) \cite{R1975,R1979} as well as for arbitrary electron concentration \mbox{$0\leq n \leq 2$} \cite{KKR2010} (these results are rigorous in the limit of infinite dimensions \mbox{$d\rightarrow+\infty$}).
Our preliminary Monte Carlo (MC) results have been presented in \cite{MKPR2012,MKPR2014} for \mbox{$L=10$} and on-site repulsion: \mbox{$U/(4J)=1,10$} (corresponding to rather weak and strong coupling, respectively)  \cite{MKPR2012} as well as for \mbox{$L=20$} and \mbox{$U/(4J)=1$} \cite{MKPR2014}.

In the present paper we investigate in details the phase diagram and thermodynamic properties of the model for arbitrary electron concentration \mbox{$n\leq1$} and arbitrary chemical potential \mbox{$\bar{\mu}\leq 0$} (\mbox{$\bar{\mu}=\mu-U/2$}) in the whole range of temperatures for \mbox{$U/(4J)=2$} and \mbox{$L=20$}. In particular, we focus on a~behavior of an~electron compressibility.
The corresponding results for \mbox{$n>1$} (\mbox{$\bar{\mu}>0$}) are obvious because of the electron-hole symmetry of the model on alternate lattices mentioned previously.

\section{Results and discussion ($U/(4J)=2$)}

Finite temperature phase diagrams for this model were obtained using MC simulations for \mbox{$U/(4J)=2$} (and \mbox{$L=20$}) as a function of $\bar{\mu}/4J$ and $n$  are presented in Fig.~\ref{fig:PhaseDiagrams}(a) and Fig.~\ref{fig:PhaseDiagrams}(b), respectively.

The behavior of this system for fixed $\bar{\mu}$ is rather simple (Fig.~\ref{fig:PhaseDiagrams}(a)), with both first-order (below $\mathbf{T}$-point) and second-order (above $\mathbf{T}$-point) phase transitions separating non-ordered (NO) and antiferromagnetic (AF) phases with tricirtical point $\mathbf{T}$ located at \mbox{$k_BT/4J=0.205\pm0.003$}, \mbox{$\bar{\mu}/4J=-1.405\pm0.007$} (\mbox{$n\simeq0.58$}).
The location of $\mathbf{T}$-point has been determined using hysteresis analysis  \cite{MKPR2014}.
In Fig.~\ref{fig:PhaseDiagrams}(a) ``heating'' and ``cooling'' label the boundaries obtained by simulation performed with increasing and decreasing temperature whereas ``average'' is the average of these two results. They differ if the \mbox{AF--NO} transition is first-order.
Details of this method can be found in \cite{MKPR2014}.

With simulations done for fixed $\bar{\mu}$ and $k_BT/4J$~vs.~$\bar{\mu}/4J$, it is  possible to obtain phase diagrams as a function of $n$ (shown in Fig.~\ref{fig:PhaseDiagrams}(b)) by determining electron density above ($n_{-}$) and below ($n_{+}$) the \mbox{AF--NO} phase transition (for fixed $\bar{\mu}$).
The first-order \mbox{AF--NO} boundary for fixed $\bar{\mu}$ splits into two boundaries (i.e. \mbox{PS--AF} and \mbox{PS--NO}) for fixed $n$.
At sufficiently low temperatures, i.e. below  $\mathbf{T}$-point, a~phase separated (PS: AF/NO) state occurs. The PS state is a~coexistence of two (AF and NO) homogeneous phases. At higher temperatures (i.e. above $\mathbf{T}$-point) the  \mbox{AF--NO} transition is second-order one.

An objective indicator of a~PS state existence is the evolution of the compressibility $K$ of the system \cite{P2006,PK2008}.
For a system with variable number of particles it can be defined as
\begin{equation}\label{row:compress}
\frac{1}{K}=n^2 \left( \frac{\partial \mu}{\partial n} \right)_{T, U, J}.
\end{equation}
From this definition it follows that at a fixed $\mu$ the number of particles in an open system can fluctuate freely (precisely, in some define range) when \mbox{$K\rightarrow\pm\infty$}.
Such a behavior is connected with  an occurrence of the PS states in define range of $n$.
At the same constant total free energy of the system  the number of domains as well as their distribution can change.
Hence, the phase separation states are ,,highly unstable'' in that sense that they are subjected to continuous fluctuations of local density (but the total density $n$ is constant).

\begin{figure}[t]
\centering
\includegraphics[width=\szerokosc\textwidth]{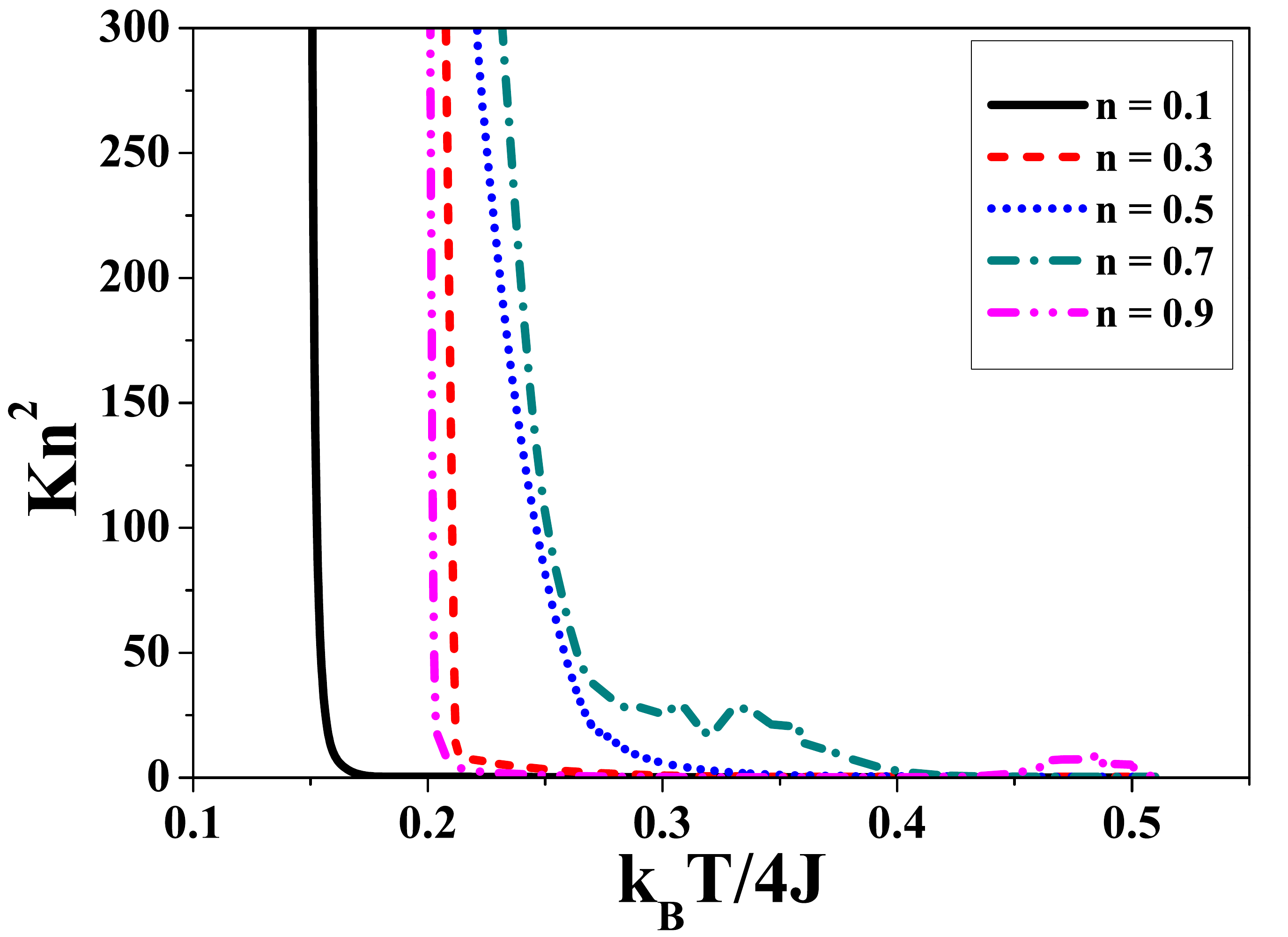}
\caption{Compressibility $K$ as a function of temperature $T$ for constant values of electron concentration $n$ (as labeled) for \mbox{$L=20$}. For \mbox{$n>0.58$} there are fluctuations of $K$ associated with second-order \mbox{AF--NO} phase transitions.}
\label{fig:ComRuns}
\end{figure}

\begin{figure}[t]
\centering
\includegraphics[width=\szerokoscdwa\textwidth]{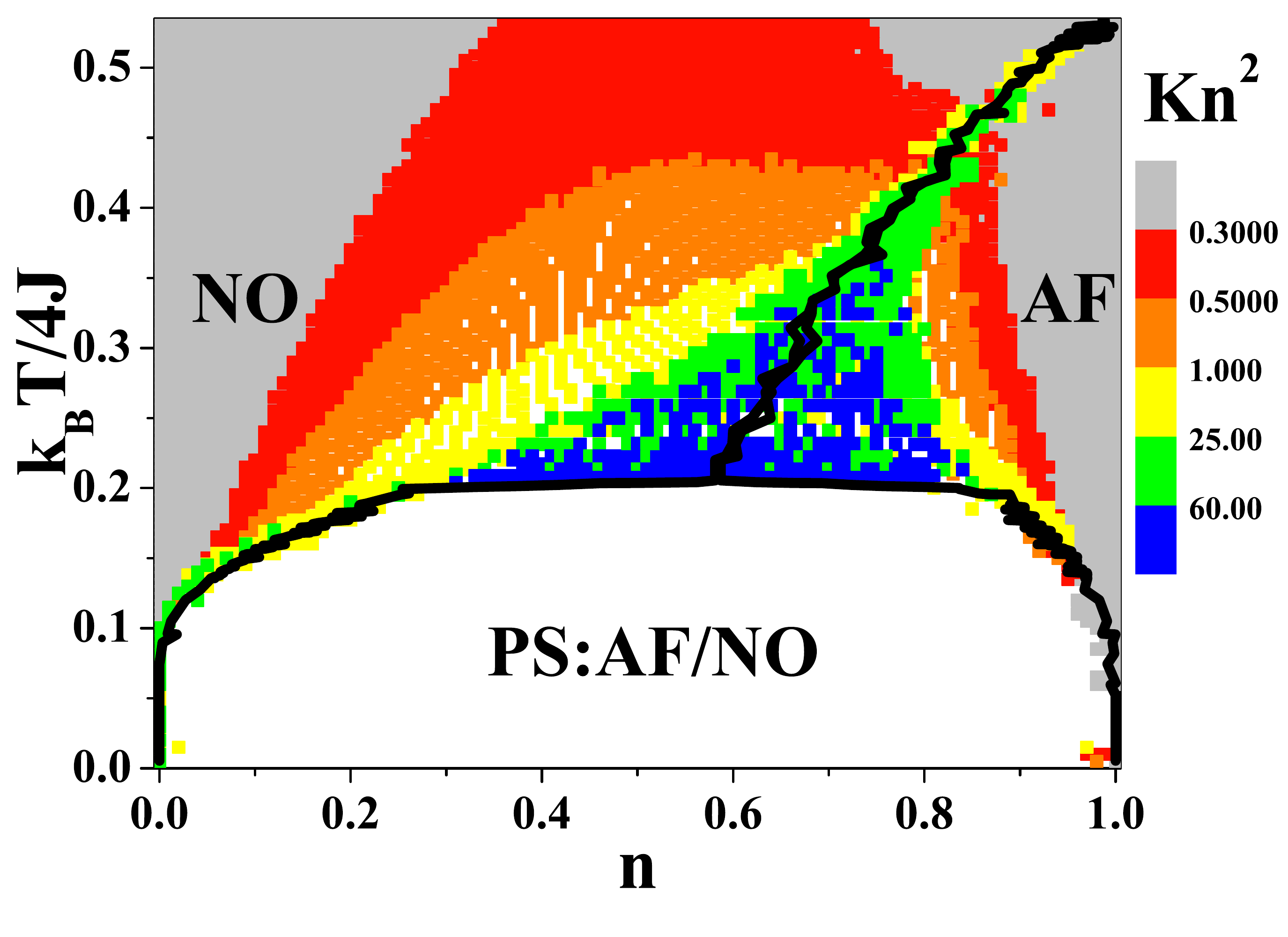}
\caption{A map of compressibility $Kn^2$ on $k_BT/4J$--$n$ plane (for \mbox{$L=20$}).
Solid black curves indicate the phase boundaries derived in Fig.~\ref{fig:PhaseDiagrams}(b) plotted for the comparison.}
\label{fig:PhaseCom}
\end{figure}

In Fig.~\ref{fig:ComRuns} the compressibility $Kn^2$ is plotted versus temperature $k_BT/4J$ for several constant values of $n$.
As is clearly seen, close to the boundaries of the PS:AF/NO state occurrence plotted in Fig~\ref{fig:PhaseDiagrams}(b), the value of compressibility abruptly increase and \mbox{$K\rightarrow+\infty$} at transition temperature, indicating an existence of the phase separation state.
At higher temperatures and for larger concentrations than those corresponding to the $\mathbf{T}$-point (\mbox{$n\gtrsim0.58$}) there are compressibility fluctuations related to second order \mbox{AF--NO} phase transition as shown on Fig.~\ref{fig:ComRuns} at higher temperatures.
They are significantly smaller than the fluctuations close to the boundaries below which  the PS:AF/NO state occurs.

Fig.~\ref{fig:PhaseCom} presents a map of compressibility on \mbox{$k_BT/4J$--$n$} plane together with the phase boundaries derived in Fig.~\ref{fig:PhaseDiagrams}(b) plotted for the comparison.
An increase of compressibility close to the second-order \mbox{AF--NO} boundary is clearly seen.
The boundary between homogeneous phases and phase separation state is also visible, with the  compressibility close to the boundary being at least an order of magnitude greater than  those inside the homogeneous phases.
As it was said earlier, in the case of phase separation occurrence $K\rightarrow\infty$, so no points  are shown inside the region of the PS state occurrence in Fig.~\ref{fig:PhaseCom}.

\section{Final comments}

Notice that the results presented are in good qualitative agreement with mean field calculations using variational approach presented in \cite{R1979,R1975,R1973,KKR2010,MKPR2012,K2012}. When comparing these results one should keep in mind differences between these two methods, as the VA is exact only for infinite dimensions $d\rightarrow\infty$.
The drawback of Monte Carlo simulations is long thermalization time, which prevents us from obtaining results for the ground state and very low temperatures, as in these conditions electrons have very small probability of escaping local energy minima.
Behavior of the model considered in the case of finite band ($t\neq 0$) is very interesting and mostly open problem in the general case \cite{MRR1990,JM2000,DJSZ2004,CzR2006,DJSTZ2006,CzR2006a}.

\begin{acknowledgments}
S.M. and K.J.K. thank the European Commission and the Ministry of Science and Higher Education (Poland) for the partial financial support from the European Social Fund---Operational Programme ``Human Capital''---POKL.04.01.01-00-133/09-00---``\textit{Proinnowacyjne kszta\l{}cenie, kompetentna kadra, absolwenci przysz\l{}o\'sci}''.
K.J.K. and S.R. thank  National Science Centre (NCN, Poland) for the financial support as a~research project under grant No. DEC-2011/01/N/ST3/00413 and as a~doctoral scholarship No. DEC-2013/08/T/ST3/00012.
K.J.K. thanks also the Foundation of Adam Mickiewicz University in Pozna\'n for the support from its scholarship programme.
\end{acknowledgments}

%%%%%%%%%%%%%%%%%%%Bibliografia%%%%%%%%%%%%%%%%

\end{document}